\documentclass[journal=nalefd,manuscript=letter]{achemso}

\usepackage{graphicx}
\usepackage{amssymb}
\usepackage{amsmath}

\author{Nojoon Myoung}
\altaffiliation{NM and HCP equally contribute to the present work}
\affiliation{Department of Material Science and Engineering,
University of Ioannina, Ioannina 45110, Greece}
\email{nmyoung@cc.uoi.gr}
\author{Hee Chul Park}
\altaffiliation{NM and HCP equally contribute to the present work}
\affiliation{Center for Theoretical Physics of Complex Systems, Institute for Basic Science, Daejeon 34051, Republic of Korea}
\author{Seung Joo Lee}
\affiliation{Quantum-functional Semiconductor Research Center,
Dongguk University, Seoul 100-715, Republic of Korea}
\email{leesj@dongguk.edu}

\title{Gate-Tunable Spin Transport and Giant Electroresistance in Ferromagnetic Graphene Vertical Heterostructures}

\keywords{Vertical Heterostructure, Ferromagnetic Graphene, Spintransport, Giant Electroresistance}

\begin{document}

\begin{tocentry}
\includegraphics[width=8.5cm]{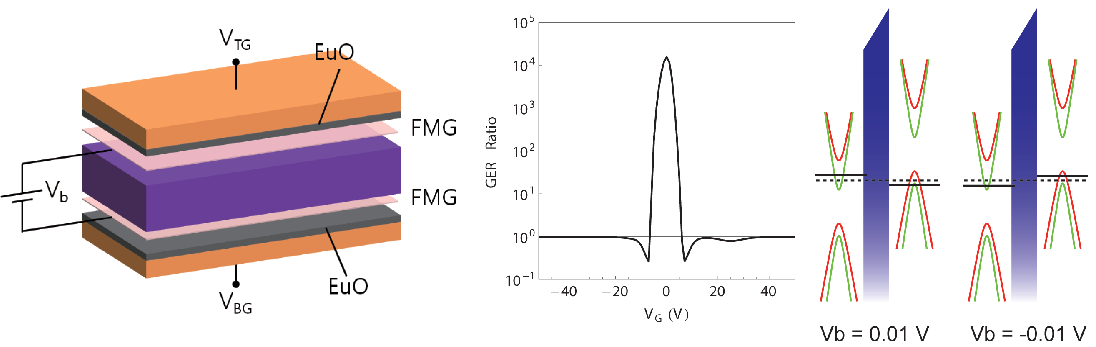}
\end{tocentry}

\begin{abstract}
We investigate spin transport through ferromagnetic graphene vertical heterostructures where a sandwiched tunneling layer is either a normal or ferroelectric insulator. We show that the spin-polarization of the tunneling current is electronically controlled via gate voltages. We also demonstrate that the tunneling current of Dirac fermions can be prohibited when the spin configuration of ferromagnetic graphene sheets is opposite. The giant electroresistance can thus be developed by using the proposed heterostructure in this study. The effects of temperature on the spin transport and the giant electroresistance ratio are also investigated. Our findings discover the prospect of manipulating the spin transport properties in vertical heterostructures through an electric fields via gate and bias electrodes.
\end{abstract}

Graphene, a honeycomb-like single layer crystal of carbon atoms, has been attracting a lot of attention in the recent decade both in terms of fundamental interests and technology. Amongst the various aspects of graphene, one of the most promising potentials is that it has extraordinary transport properties such as high carrier mobility and long mean free path\cite{Novoselov2004,Bolotin2008a,Bolotin2008b}. Despite these advantages for high-speed device applications, the use of single layer graphene for practical nanoelectronic devices, like field-effect transistors (FETs), is limited because of the low current on/off ratio\cite{Lemme2007,Meric2008,Kim2009} that implies how effectively it generates digital signals. This limitation mainly stems from the intriguing relativistic transport phenomena in graphene, so-called Klein tunneling, which results in massless and chiral Dirac fermions that can perfectly pass through electrostatic potential barriers\cite{Katsnelson2006,Beenakker2008,Stander2009}.

Recently, there has been an alternative idea to fabricate graphene FETs based on an architecture that graphene and other two dimensional layers are stacked vertically\cite{Britnell2012a,Georgiou2013}. It has been reported that vertical current density in layered graphene - hexagonal boron nitride (hBN) -graphene heterostructures can be modulated by controlling quantum tunneling through atomically thin hBN layers via gate voltage\cite{Britnell2012a,Fiori2013}. Larger current on/off ratios can be achieved by using small-bandgap layered materials as a tunneling insulator\cite{Britnell2012a,Myoung2013}. Owing to the huge variety of structures and properties in vertical heterostructures of 2D materials, many promising and interesting research topics have been considered, e.g., field-effect transistors\cite{Britnell2012a,Britnell2012b}, resonant tunnel diodes\cite{Britnell2013a,Mishchenko2014}, and photodetectors\cite{Britnell2013b,Yu2013}. In particular, the vertical heterostructure architecture can be also a good candidate for graphene-based spintronics when sandwiched insulating layers are magnetized\cite{Myoung2013}. The long spin-coherent length of graphene\cite{Huertas-Hernando2007,Avsar2011,Kozikov2012} allows for the fabrication of spintronic devices using graphene sheets as spin transport channels, once the tunneling current is well spin-polarized through the vertical heterostructure. 

In this letter, the spin-resolved transport through the vertical heterostructures with ferromagnetic graphene (FMG) is investigated. Also, it is shown that the control of the spin transport through the structure can be achieved by electrically manipulating the spin configurations in FMG sheets.  The spin-resolved band structure is taken into account to describe the electronic states of FMG, and the  spin-resolved tunneling current density is calculated for two different combinations of heterostructures: FMG - normal insulator(NI) - FMG and FMG - ferroelectric insulator (FEI) - FMG. It is shown that the giant electroresistance emerges for the anti-parallel configuration of FMGs when the sandwiched insulator is replaced by an FEI.

\begin{figure}
\includegraphics[width=8.4cm]{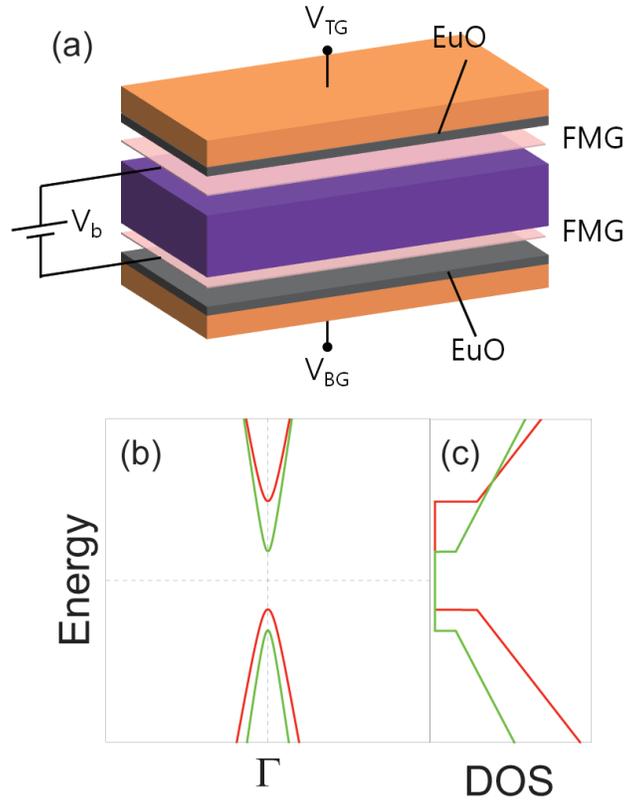}
\caption{Model of heterostructure and electronic properties of ferromagnetic graphene (FMG). (a) Schematics of the vertical heterostructures with FMG and a tunneling insulator. (b) Spin-resolved band structures and (c) spin density of states (SDOS) of FMG. Red and Green solid lines represent spin-up and down states in FMG with spin-resolved bandgaps and Fermi velocities.} \label{fg:model}
\end{figure}

The system studied in this letter is a vertically stacked heterostructure which is formed by FMG and an insulating layer [see Figure \ref{fg:model}a]. The sandwiched insulator and the graphene sheets play roles of a tunnel barrier and conducting channels, respectively. Dual-gated device structures are considered to control the same amount of the carrier densities on both graphene sheets\cite{Rodriquez-Nieva2015}. It has been revealed that the proximity interaction between a ferromagnetic insulator such as europium oxide (EuO) and graphene is able to induce ferromagnetism in graphene\cite{Swartz2012,Yang2013,Wang2015,Qiao2014}. The bias voltage $V_{b}$ can be applied between two graphene sheets, yelding the tunneling current through the insulating layer. The electronic properties of FMG are characterized by its spin-resolved electronic states\cite{Yang2013}:
\begin{align}
E_{\sigma}\left(q\right)=\pm\sqrt{\left(\hbar v_{\sigma}q\right)^{2}+\left(\frac{\Delta_{\sigma}}{2}\right)^{2}},
\end{align}
where $\sigma=\pm1$ for spin-up and down states of Dirac fermions, $v_{\uparrow}=1.15\times v_{F}$ and $v_{\downarrow}=1.4\times v_{F}$ are Fermi velocities for each spin with $v_{F}=10^{6}$ m/s, $\Delta_{\uparrow}=134$ meV and $\Delta_{\downarrow}=98$ meV are the spin-resolved bandgaps, as displayed in Fig. \ref{fg:model}(b). Here, we assume that the Fermi level of FMG is set in the mid-gap. For the proximity-induced ferromagnetic graphene, the valley degeneracy of the pristine graphene has been broken by interactions between carbon and europium atoms\cite{Yang2013}. Particularly, depending on the position of the Fermi level, the FMG can be fully spin-polarized - at positive or negative unity - by adjusting the gate voltage via both gate electrodes (see Supporting Information).

As a starting point, let us introduce our vertical transport model used in this study. It assumes the elastic tunneling of Dirac fermions in terms of energy, and the momentum scattering effects are taken into account by applying the current density formula. The spin-resolved vertical tunnelig current is formulated with the interlayer transition matrix element based on WKB approximation,\cite{Rodriquez-Nieva2015}
\begin{align}
j_{\sigma}=\frac{e}{h}\int \left|t\left(\epsilon\right)\right|^{2}D_{1,\sigma}\left(\epsilon+\frac{eV_{b}}{2}\right)D_{2,\sigma}\left(\epsilon-\frac{eV_{b}}{2}\right)\left[f_{1}\left(\epsilon+\frac{eV_{b}}{2}\right)-f_{2}\left(\epsilon-\frac{eV_{b}}{2}\right)\right]d\epsilon, \label{eq:curden}
\end{align}
where $D_{\sigma}\left(\epsilon\right)=\left|\epsilon\right|/\left(2\pi\hbar^{2}v_{\sigma}^{2}\right)\Theta\left(\left|\epsilon\right|-\Delta_{\sigma}/2\right)$ is the spin-resolved density of states (SDOS) with the spin-resolved Fermi velocities (see Figure \ref{fg:model}c), and $f\left(\epsilon\right)=\left[1+e^{-\left(\epsilon-\mu\right)/\left(k_{B}T\right)}\right]^{-1}$ is the Fermi-Dirac distribution. The interlayer transition matrix element is given by
\begin{align}
t\left(\epsilon\right)=\Gamma e^{-\left(1/\hbar\right)\int_{-d/2}^{d/2}\sqrt{2m^{\ast}\left(\Delta-eV_{b}z/d-\epsilon\right)}dz}. \label{eq:transprob}
\end{align}
Here, $m^{\ast}$ is the effective mass of the tunnel barrier material, $\Delta$ is the barrier height of the tunneling insulator, $V_{b}$ is bias voltage which is applied via two graphene sheets, and $d$ is the thickness of the tunneling insulator. Note that $\Gamma$ is an energy-independent prefactor which represents the momentum scattering of Dirac fermions by disorders such as defects or phonons inside the tunnel barrier material. In other words, for $\Gamma=1$, there is no scattering mechanism while Dirac fermions tunnel through the tunnel barrier, and on the other hand, the smaller $\Gamma$s indicate more diffusive vertical transport through the tunnel barrier.

Carrier density on graphene layers is controlled by field-effects via gate electrodes. In the absence of bias voltage, the chemical potentials on both graphene layers are in equilibrium, leading to no net tunneling current density. The dual-gated platform is considered to fix and maintain the same carrier densities in the top and the bottom gate electrodes\cite{Rodriquez-Nieva2015}, resulting in a symmetric gated structure. This assumption allows us to simplify the problem with fixed chemical potentials on both graphene layers in equilibrium, i.e., $\mu_{0}=\hbar^{2}v_{F}^{2}\sqrt{4\pi\left|n_{0}\right|}$.  By using the electrostatic capacitor model, $n_{0}$ is proportional to the gate voltage $V_{G}$, i.e., $n_{0}=\alpha V_{G}$ where $\alpha$ is the proportional constant depending on the substrate (superstrate) materials between a graphene layer and the bottom (top) gate electrode. When bias voltage is applied to both graphene layers, their chemical potentials are shifted and equilibrium is broken, resulting in non-zero tunneling current denslty throughout the vertical heterostructure.

\begin{figure}
\includegraphics[width=16.0cm]{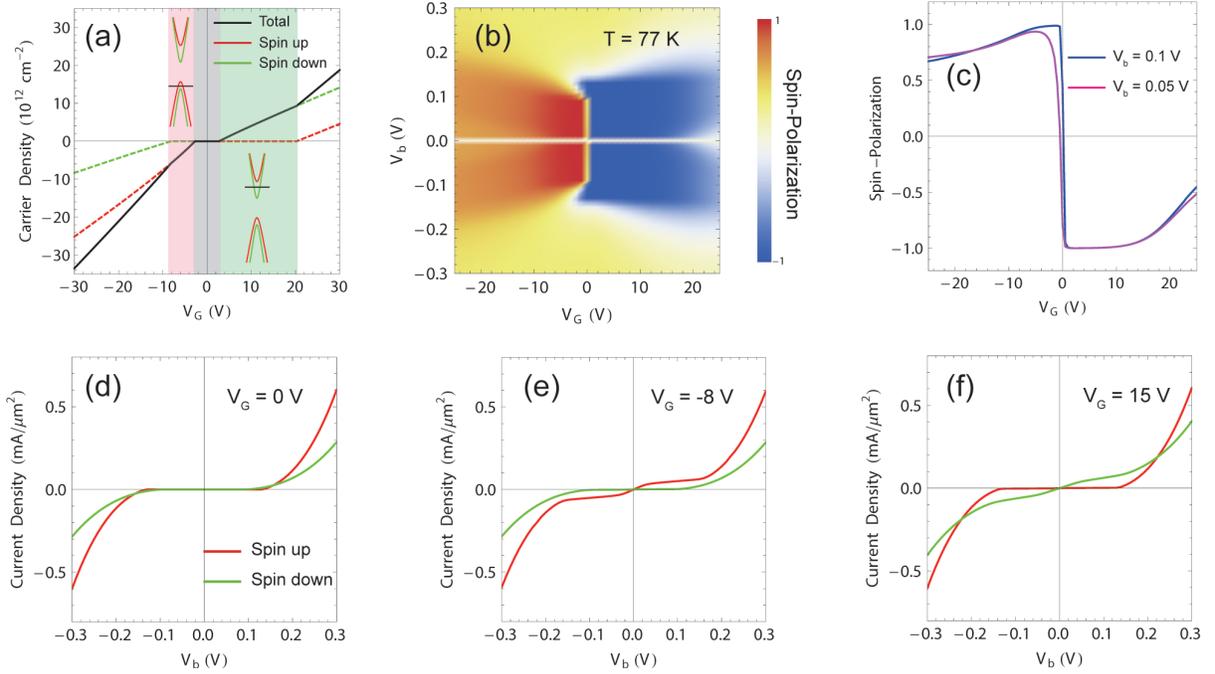}
\caption{ Spin transport through an FMG-NI-FMG hetrostructure. (a) Spin-resolved carrier densities and the corresponding total carrier density on a FMG layer versus gate voltage, in the absence of bias voltage. Left and right shaded regions represent  pure spin-polarization, which are denoted as inset diagram. Middle shaded region corresponds to the forbidden zone where no Dirac fermions are allowed. (b) Color map of the spin-polarization of the tunneling current density as functions of bias and gate voltages. For $V_{b}=0$ V, the spin polarization is defined as zero since there is no tunneling current regardless of gate voltage. (c) Spin-polarization of the tunneling current densities as functions of gate voltage for different bias voltages $V_{b}=50$ and 100 mV. (d-f) Spin-resolved tunneling current densities as functions of the bias voltage for different gate voltages $V_{G}=0$, -8, and 15 V, respectively. The results are calculated at $T=77$ K.} \label{fg:FMG-NI-FMG}
\end{figure}

Figure \ref{fg:FMG-NI-FMG} shows spin-resolved vertical transport through the FMG-NI-FMG heterostructure. In the present study, $\Delta=1.5$ eV and $m^{\ast}=0.5~m_{el}$ with the bare mass of an electron$m_{el}$ are used for the calculations, which are approximately compatible with typical 2d materials such as MoS$_{2}$, WS$_{2}$, etc.\cite{Mak2010,Huang2015,Kuc2011,Braga2012} In the same context, the thickness of the NI layer is taken as 1 nm which is compatible with few-layer 2d material cases\cite{Britnell2012b}. The spin transport is characterized by the spin-polarization of the tunneling current density,
\begin{align}
P_{j}=\frac{j_{\uparrow}-j_{\downarrow}}{j_{\uparrow}-j_{\downarrow}}.
\end{align}
As Figure \ref{fg:FMG-NI-FMG}b exhibited, the tunneling current density is well spin-polarized for small amounts of bias voltage, $\left|V_{b}\right|<0.1$ V. Remarkably, within this bias voltage range, it is found that the current density can be fully spin-polarized according to gate voltage. This gate-tunable feature of the spin transport is led by the following mechanisms. When the equilibrium chemical potential $\mu_{0}$ is place in the mid-gap, the small bias voltage cannot lead to a sufficient amount of tunneling current densities for both spins. As bias voltage increases, spin-resolved current densities begin to flow. Here, one can see that the spin-down current starts flowing slightly earlier than the spin-up current density because of their different electronic properties, i.e., the amount of band gaps and the position of the band edges. For large bias voltages, the current densities for both spins keep increasing with different increasing ratios associated to the spin-resolved Fermi velocities in SDOS. For $V_{G}=15$ V, as plotted in Figure \ref{fg:FMG-NI-FMG}a, an FMG layer is purely spin-down-polarized, and thus a pure spin-down current is generated by small bias voltages (see Figure\ref{fg:FMG-NI-FMG}f). On the contrary, an application of $V_{G}=-8$ V makes an FMG layer purely spin-up polarized, and the contribution to the tunneling current density is dominated by spin-up Dirac fermions for small bias voltages as shown in Figure\ref{fg:FMG-NI-FMG}e. In other words, the spin-polarization of the tunneling current density can be  switched according to the gate voltage, as shown in Figure \ref{fg:FMG-NI-FMG}c. There is a very large contrast in the spin-polarization values around $V_{G}=0$ V because the majority spin states near both band edges are opposite to each other (see Figure\ref{fg:model}c). Note that the spin transport phenomena are influenced by temperature, but this spin-switching effects are expected to be observed even at room temperature (see Supporting Information). Besides, one can see that the spin-up contribution to the tunneling current is always dominant for the relatively larger bias voltages in Figures \ref{fg:FMG-NI-FMG}d-f. This results from the fact that an FMG ends up spin-up-polarized as its Fermi level is tuned away from the band gap (see Figure \ref{fg:model}c).

For FMG-NI-FMG heterostructures, it is found that the spin-resolved band structure of FMG is involved in the spin-polarized tunneling phenomena and the manipulation of the spin degree of freedom by means of an electric field via gate electrodes. The occurrence of pure spin-polarized current is attributed to the spin-resolved band gap of FMG, where only one spin states can be allowed near the band edges. This feature leads to purely spin-polarized FMG layers which can be utilized in the spintronic devices to explore a giant magnetoresistance (GMR). For typical ferromagnetic metal (FM)-NI-FM heterojunctions, electrical resistance strongly depends on how the FM configuration is set. While the electrical current flows well with the small resistance in the parallel configuration, a very large resistance is measured in the anti-parallel configuration. To achieve GMR, devices should be asymmetrically fabricated by using different kinds of FM materials, for which magnetization varies with respect to external magnetic fields. This means that controlling the magnetic fields is essential to change FM configuration. In addition, GER has already been introduced in a normal metal (NM)-FEI-NM heterojunction by using the asymmetric electrical response of a sandwiched FEI\cite{Zhuravlev2005}. The key to GER is using electric fields instead of magnetic fields to achieve a giant change in electrical resistance, allowing greater convenience in generating distinct on/off signals in terms of technology. However, an asymmetric device has still been essential to make the potential barrier profile inside the FEI layer. Here, a way of achieving the emergence of GER is presented by investigating vertical transport through FMG-FEI-FMG vertical heterostructures. Our device architecture not only has an ability to produce a giant resistance change by means of electric fields, but also does not require asymmetric fabrication.

The properties of an FEI are described by a simple model of the polarization density as a function of an external electric field,
\begin{align}
P\left(E_{b}\right)=p_{0}\tanh{\left[\beta\left(E_{b}-s\cdot E_{c}\right)\right]},
\end{align}
where $p_{0}$ is the saturated polarization density, $\beta$ is the characteristic coefficient with the physical dimension of inverse electric fields, $\vec{E}_{b}=-V_{b}/d$ is an external electric field applied via bias voltage, and $E_{c}$ is the coercive field which is responsible for the hysterisis of the FEI. Here, the factor $s=\pm1$ implies how the electric field varies, i.e. the forward or reverse sweep of electric fields (see Supporting Information for the hysterisis of an FEI). The presence of the ferroelectricity in the tunnel barrier material is reflected two-fold. i) Carrier density on the FMG layers are influenced by the bound charge at FEI interfaces, $\sigma_{b}=\vec{P}\cdot\hat{n}$. Accordingly, the spin-resolved Dirac cones are shifted by the amount of the charge imbalance between the FMGs. ii) There is an additional tunnel barrier induced by the bound charges, besides the tunnel barrier caused by an external field.  The former offers a rearrangement of the spin-resolved Dirac cones on FMG layers, and the latter accounts for the direction-dependent tunneling probability of Dirac fermions.

The shift of the spin-resolved Dirac cones is led by the following mechanism. For dual-gated devices, the carrier densities on FMG layers $n_{1}$ and $n_{2}$ are given as $n_{1}=n_{0}-\delta n/2$ and $n_{2}=n_{0}+\delta n/2$, where $\delta n=\sigma_{b}$. The corresponding chemical potentials are determined by $n_{1}$ and $n_{2}$, i.e., $\mu_{1,2}=sgn\left(n_{1,2}\right)\sqrt{\pi\left|n_{1,2}\right|}$, where $sgn\left(n_{1,2}\right)$ is the sign function. In equilibrium, the chemical potentials on the FMG layers should be arranged at the same Fermi energy to be consistent with equilibrium in the absence of bias voltage. Therefore, the Dirac cone on each FMG layer is shifted by $\pm\delta\mu/2=\pm\left|\mu_{1}-\mu_{2}\right|/2$, respectively. In fact, such Dirac cone shifts coincide with a uniform electric field inside the tunneling layer $\delta\mu/ed$. This FEI-induced electric field is reflected in the tunnling probability as below,
\begin{align}
t\left(\epsilon\right)=\Gamma e^{-\left(1/\hbar\right)\int_{-d/2}^{d/2}\sqrt{2m^{\ast}\left(\Delta-eV_{b}z/d-\delta\mu z/d-\epsilon\right)}dz}.
\end{align}

\begin{figure}
\includegraphics[width=16.0cm]{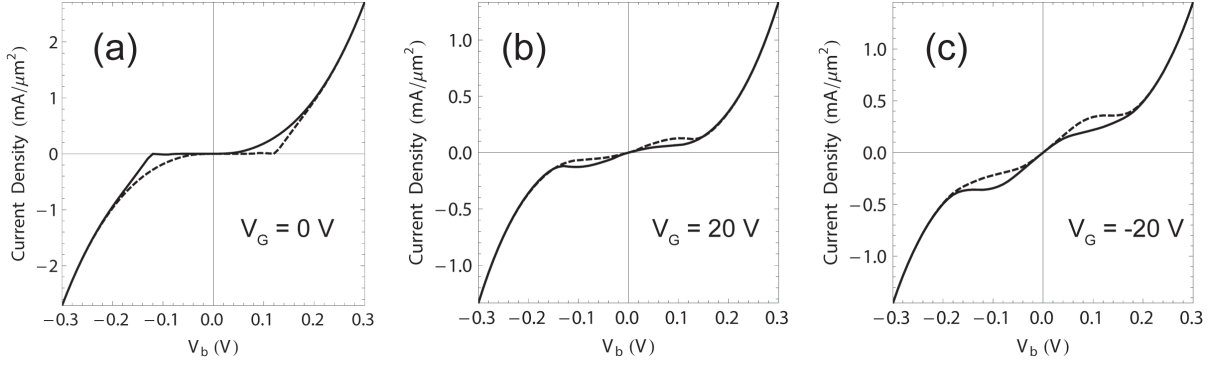}
\caption{ Total tunneling current densities versus bias voltage for different gate voltage in an FMG-FEI-FMG heterostructure. Solid and dashed lines represent the forward and reverse sweeps of bias voltage. } \label{fg:FEcurrent}
\end{figure}

Figure \ref{fg:FEcurrent} presents the vertical transport properties through FMG-FE-FMG heterostructures. Here, total current density is shown as a function of bias voltage, which is given by the sum of the spin-up and spin-down current densities. It is clearly shown that the tunneling current density exhibits hysterisis behavior associated with FEI nature. For large bias voltage, the current density with the forward bias sweep is the same as that with the reverse bias sweep, resulting from the saturation of the polarization density. Total current density is resolved into spin-up and down current densities, and the spin-resolved feature is helpful in understanding the sweep-direction dependence (see Supporting Information). Also, it is found that there is a considerable dependence of the current density behavior on gate voltage.

\begin{figure}
\includegraphics[width=16.0cm]{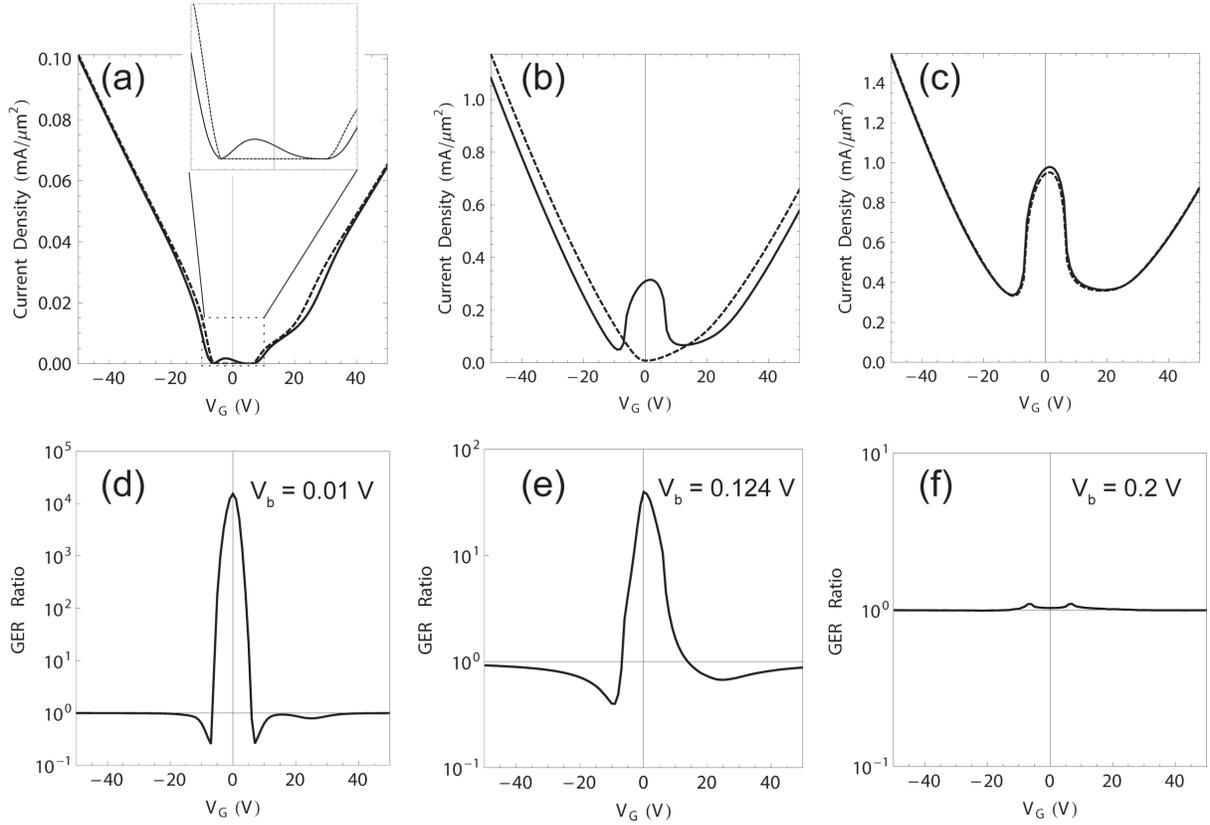}
\caption{Giant electroresistance (GER) of FMG-FEI-FMG heterostructures. (a) Tunneling current densities versus gate voltage for $V_{b}=\pm0.01$ V. Inset: Close-up of the current density plots for different bias voltage directions. (b,c) Tunneling current densities versus gate voltage for $V_{b}=\pm0.124$ and $\pm0.2$ V. (d,e,f) GER ratios as functions of gate voltage for different magnitudes of bias voltages, which correspond to (a,b,c), respectively. Solid and dashed lines represent the positive and the negative bias voltage. The results shown here are calculated at $T=77$ K and for the forward sweep direction.} \label{fg:GERratio}
\end{figure}

\begin{figure}
\includegraphics[width=15.0cm]{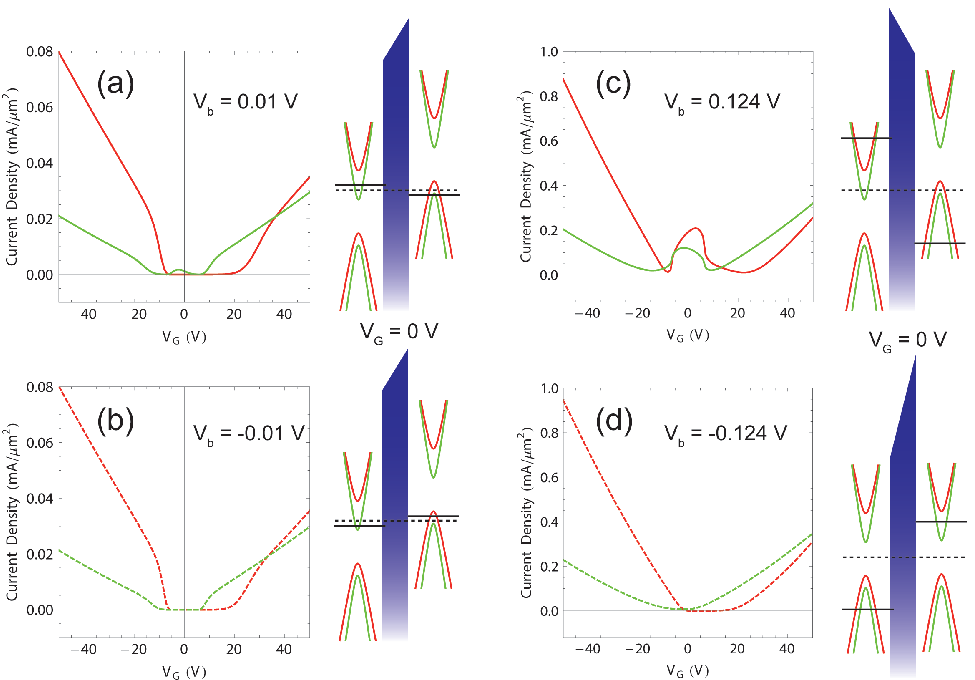}
\caption{ Spin-resolved vertical transport through FEI tunnel barriers for different bias voltage directions. (a,b) Plots of the tunnleing current densities attributed to different spins as functions of gate voltage, for $V_{b}=\pm0.01$ V, respectively. (c,d) Plots of the tunnleing current densities attributed to different spins as functions of gate voltage, for $V_{b}=\pm0.124$ V, respectively. Energetic diagrams next to each plot present the corresponding interpretations of the tunneling mechanism where the shift of the spin-resolved Dirac cones and the positions of the chemical potentials on the FMG layers. Absolute values of the current density are shown for the both positive (solid lines) and negative (dashed lines) bias voltages.} \label{fg:GERspintrans}
\end{figure}

Due to the hyeterisis feature, the current density values are expected to be asymmetric with respect to the bias voltage polarity, for a sweep direction of bias. Such an asymmetric response to bias voltage makes the current density $j_{+}^{for,rev}$ for positive bias voltage different from the current density $j_{-}^{for,rev}$ for negative bias voltage, thereby resulting in a large ratio between them. Here, let us define GER ratio as $\left|j_{+}^{for}/j_{-}^{for}\right|$ for the forward sweep direction and $\left|j_{-}^{rev}/j_{+}^{rev}\right|$ for the reverse sweep direction. As shown in Figure \ref{fg:GERratio}g-f, the GER ratio converses to unity as gate voltage increase because of the fact that the FEI-induced shift of the Dirac cones cannot result in considerable differences in the FMG configuration. For convenience of comparison, the absolute values of the current densities are displayed. In general, the largest GER ratios are found around $V_{g}=0$ V where chemical potentials reside near the band edges of FMGs. Further, for the very small bias $V_{b}=0.01$ V, the tunneling current is allowed only by a positive bias, whereas it is strongly suppressed by a negative bias. Such a large GER ratio is led by the following mechanism. For very small bias voltages, the polarization density is almost unchanged from the saturated value, and the resulting Dirac cone shift makes one FMG layer purely spin-up polarized and the other FMG layer purely spin-down polarized, i.e., anti-parallel spin configuration of FMGs is derived. When a small bias is applied in the positive direction, the chemical potential on spin-up polarized FMG becomes lower and touches the lower spin-down band, while the chemical potential on spin-down polarized FMG becomes higher but still reside in the spin-down band only. On the other hand, the FMG configuration remains anti-parallel for $V_{b}=-0.01$ V, resulting in the suppression of vertical tunneling by Pauli blocking. Indeed, Figure \ref{fg:GERspintrans}a shows that the current density for $V_{b}=0.01$ V is influenced by the spin-down states only. Therefore, the GER ratio $\sim10^{4}$ originates from the bias-tunable spin-configuration of FMGs. Also, the GER ratio has dependence on temperature and deteriorates at higher temperatures (see Supporting Information).

The effects of the FEI-induced Dirac cone shift are well interpreted in Figures \ref{fg:GERspintrans}b,e. In this case, the applied bias is associated with the coercive fields, which make the polarization density of an FEI according to the sweep direction of bias voltage. For a forwardly sweeping bias, it is found that $V_{b}=-0.124$ V leads to zero polarization density, and $V_{b}=+0.124$ V makes the polarization density saturated. In other words, for $V_{b}=-0.124$ V, no shift is induced between two FMG layers, mimicking an FMG-NI-FMG heterostructure. Indeed, the tunneling current density exhibits behavior of typical vertical FETs where the tunneling current through an insulating layer is controlled by gate voltage. When the bias voltage is reversed to $+0.124$ V, the spin-resolved Dirac cones are shifted by the saturated polarization density of the FEI, and the tunneling current begins to flow even for zero gate voltage (see Figures \ref{fg:GERspintrans}c,d). Due to the relatively large bias voltage, the energy window is wide enough to allow both spin-up and spin-down tunneling currents. In this case, the tunneling current density exhibits a distinct behavior as gate voltage increases: the current density drops for specific gate voltage because the chemical potential of one FMG layer falls into a band gap, and then both chemical potentials reside in upper (or lower) bands, making the current density increases again as gate voltage increases.

In summary, it is demonstrated that the tunneling current density can be spin-polarized through FMG-NI-FMG heterostructures, reaching up to unity. By using the spin-resolved band model of the FMG, it is revealed that the vertical transport is accordingly spin-resolved. The spin transport through the FMG-NI-FMG heterostructure depends on the position of the equilibrium chemical potential, and its spin-polarization of the current density is tunable via gate voltage. This gate-tunable spin transport is attributed to the presence of the purely spin-polarized states in the FMG band model, which can be a good building block for GMR devices. Accordingly, it is also demonstrated that the FMG heterostructure can be utilized to generate GER by replacing an NI with an FEI. Due to the FEI-induced shift of the FMG bands, it is shown that the anti-parallel spin configuration is achieved for specific gate voltages, and the spin configuration is able to be manipulated by means of electric fields via bias voltage. For specific gate voltages applied to the system, a very large difference in the tunneling current density is observed according to the bias voltage polarity (positive or negative). The influence of temperature on the spin-polarized tunneling current and the GER ratio is also investigated. As temperature increases, both the gate tunability of the spin-polarization and the GER ratio deteriorate, but the fully spin-polarized tunneling current and the very high GER ratio are guaranteed at $T=77$ K. Gate-tunable spin transport can present a new means of manipulation of spin states using electric fields rather than magnetic fields, based on vertical heteroctructure architectures.

\begin{suppinfo}
Model of a ferroelectricity considered in FMG/FEI/FMG heterostructure for large GER ratios, and temperature dependence of both spin-polarized tunneling current and GER ratios. This materials is available free of charge via the Internet at http://pubs.acs.org.
\end{suppinfo}

\begin{acknowledgement}
The research leading to these results has received funding from the European Union Seventh Framework Programme under grant agreement n$^{\circ}$604391 Graphene Flagship, Project Code (IBS-R024-D1), and the NRF grant funded by MSIP(No. 2014-066298).
\end{acknowledgement}

\bibliography{FMG_VFET}

\end{document}